**Enhancing Physics Hand-on Lab through Online Educational Tools**

Marina Babayeva*

Department of Physics Education
Charles University
Faculty of Mathematics and Physics
V Holešovičkách 747/2 (7th floor)
180 00 Praha 8
Czech Republic

E-mail: marina.babayeva@matfyz.cuni.cz
ORCID ID: 0000-0002-5778-9722

**Abstract**

The increasing availability of digital tools for education offers significant opportunities to enhance teaching practices and student engagement. This study presents a structured categorization of online educational tools based on their core functionalities, including content creation, assessment, classroom management, and collaboration. A pilot implementation of selected tools was conducted in secondary-level science education, followed by a refinement phase to address usability and integration challenges. Feedback from participating teachers and workshop attendees highlighted the importance of accessibility, intuitive interfaces, and support materials. Observations revealed that while multiple tools offer broad functionality, unified platforms may better support effective instruction. The resulting categories and experiences provide practical guidance for educators seeking to integrate digital tools into their teaching in a purposeful and inclusive way.

**Keywords:** physics education, educational technology, digital tools in education, online learning platforms, instructional software

# 1. Introduction

The COVID-19 pandemic necessitated an abrupt and large-scale transition from traditional face-to-face instruction to fully online education across schools and universities worldwide. This shift significantly impacted instructional strategies, especially in disciplines requiring hands-on engagement, such as physics. A bibliometric study by Jatmiko et al. (2021) identified "online learning" as the second most prominent research trend in physics education during the pandemic period, following "experiments", highlighting the central role that digital platforms assumed in response to these unprecedented challenges.

While the immediate need for purely online instruction has since diminished, educational practices continue to evolve toward hybrid or blended learning models. These approaches combine the flexibility and accessibility of online education with the interactive and practical benefits of in-person instruction. Studies by Guo et al. (2023) and Xu et al. (2023) have demonstrated that blended learning can enhance student outcomes, especially when effectively integrated with curriculum goals and supported by appropriate technologies. Blended models enable continued access to recorded lectures, diverse digital materials, and asynchronous collaboration while preserving opportunities for hands-on experimentation, face-to-face discussion, and real-time feedback.

However, the integration of such technologies into physics education presents both pedagogical opportunities and implementation challenges. Instructors must adapt to new instructional roles, navigate steep learning curves associated with unfamiliar tools, and ensure inclusive and equitable access for all learners. Consequently, research interest in educational technology within physics education has expanded significantly in recent years. A bibliometric analysis by Prahani et al. (2022) confirmed a marked increase in the number of publications in this area during 2020–2021, reflecting the field's growing relevance and dynamism.

The educational technology market is highly diverse, with platforms tailored to different educational levels and instructional goals. Interfaces designed for primary education often prioritize simplicity and visual appeal, while platforms targeting secondary and post-secondary institutions tend to offer more complex functionality, such as advanced assessment tools, collaborative environments, and detailed performance analytics. Some tools are built for specific purposes, such as quiz creation, class management, or instructional content delivery, while others aim to integrate several functionalities within a single ecosystem. In this work, I examine several functional categories of educational software and illustrate their implementation with practical examples from classroom settings.

The increasing pervasiveness of technology in everyday life has encouraged educators to integrate digital tools into their instructional practices. Motivations for doing so include enhancing student engagement, supporting interactive and differentiated instruction, and equipping learners with digital competencies often referred to as 21st-century skills. Aprilo et al. (2023) conducted a comprehensive literature review on the integration of technology in 21st-century physical education, underlining the broader relevance of such technologies across various disciplines, including science education.

Educational technologies can also serve as scaffolds for the development of both cognitive and collaborative skills. For example, Schanze, Groß, and Hundertmark (2020) investigated the use of online tools to support collaborative writing in educational contexts. Their findings suggest that, while individual contributions remain important, technology-enhanced collaboration fosters a stronger sense of group identity and shared goals. Similarly, Salas-Rueda et al. (2022) examined the implementation of collaborative digital walls in physics education and found that such tools contributed positively to the teaching–learning process by encouraging student interaction and active knowledge construction. Collectively, these studies reflect a growing interest in the pedagogical potential of learning management software (LMS) and related technologies across educational contexts. As digital tools continue to evolve, their thoughtful integration into the science classroom offers promising opportunities to enrich both teaching practices and student learning outcomes.

The advantages and disadvantages of educational software in science education at schools and colleges have been extensively discussed in the literature. The use of educational technology offers enhanced understanding of the material being studied, inclusive teaching, increased accessibility, and providing immediate feedback (Fonseca et al., 2013). Additionally, the integration of technology in education allows for the facilitation of the learning process, as it can be used both inside and outside the classroom, thereby promoting auto-regulation of learning (Cacabelos et al., 2015). Simanullang et al. discuss the application of Moodle LMS in physics education. Moodle was chosen based on their previous research on available open-source LMS as one of the most popular systems. It offers video-based activities, forums, materials, and quizzes. According to Simanullang, students were able to successfully conduct all the activities without any obstructions (Simanullang & Rajagukguk, 2020). Setiawan et al. also talk about the LMS and use iSpring Free for creating engaging and responsive courses. They particularly underline that providing students with online materials increased accessibility of the materials anytime and anywhere given the existing internet connection (Setiawan et al., 2022). Rizal et al. discuss the development of LMS for pre-service teachers to increase their digital literacy. Overall, it is stated that the developed LMS was beneficial for students to stimulate them to develop new skills. One of the limitations was the difficulty in the case of limited/bad internet connection (Rizal et al., 2022).

Another noteworthy example of educational software in physics instruction is presented by Solvang and Haglund (2021), who investigate the use of GeoGebra, a dynamic mathematics software, in physics education. Their study offers a broad range of application scenarios across different physics domains such as mechanics, wave phenomena, and geometrical optics. By compiling and analyzing existing implementations, the authors demonstrate that the integration of GeoGebra can significantly enhance students' conceptual understanding, supporting more interactive and visual forms of learning.

Despite the promising benefits of educational technologies, their implementation is not without challenges. One such issue pertains to asynchronous learning environments, which may limit immediate feedback and student-instructor interaction. Levin (2023) notes that such scenarios necessitate advanced data analysis techniques, such as cluster analysis, to identify and address learning deficiencies effectively. Furthermore, the cognitive demands associated with digital learning environments are a matter of ongoing debate. Skulmowski and Xu (2021) examine factors contributing to cognitive load in digital education, identifying five critical elements: interactivity, immersion, disfluency, realism, and the presence of redundant information. These factors can variably influence learners' cognitive processing and educational outcomes.

The relationship between digital interactivity and learning outcomes remains inconclusive. For instance, Schubertová et al. (2023) report that although digital media offer enhanced interactivity, this does not consistently translate into improved academic performance. Additionally, concerns about students' attention spans have become increasingly prominent in discussions of blended and distance learning (Levitin, 2015; Suzuki, 2015). As students spend more time engaging with digital content, they are exposed to frequent distractions, particularly from social media platforms, which can disrupt sustained focus and diminish memory retention (Newport, 2019). Tripathi (2023) argues that the pervasive use of social media negatively impacts students' ability to concentrate, thereby undermining the learning process.

The integration of technology into education necessitates a thoughtful approach, particularly in addressing potential limitations such as disparities in student engagement and comprehension across synchronous and asynchronous learning environments (Levin, 2023). While technology in science education offers considerable advantages, including improved conceptual understanding, personalized learning pathways, and greater flexibility in instructional delivery, they also introduce notable challenges. These include difficulties in maintaining effective communication in asynchronous contexts, complexities associated with managing distance learning, and the demand for robust systems to monitor and analyze student performance data. Consequently, successful implementation of educational technologies requires not only technical infrastructure but also pedagogical strategies tailored to diverse learning needs and environments.

In this paper, the implementation and practical application of various online educational platforms in the context of physics education is examined. The study presents findings from a pilot classroom study and two educator workshops, exploring both the benefits and challenges associated with different tools. The study aims to offer a categorized overview of currently available digital resources, identify their pedagogical potential, and provide guidance for educators seeking to enrich their teaching through the thoughtful integration of online platforms.

## 2. Method

Initial Tools Search and Categorization

This study constitutes an initial phase of a broader research project investigating the role of digital tools in supporting hands-on physics laboratories. To inform the development of instructional materials, I conducted a preliminary review of freely available educational technologies that support three core functions: access to learning content, opportunities for problem-solving practice, and mechanisms for formative and summative assessment.

The search strategy was designed to reflect the typical behavior of educators seeking digital tools. It included:

- Keyword-based online searches using general-purpose search engines (e.g., Google);
- Review of widely used educational video tutorials;
- Analysis of popular thematic blog posts focused on technology in science education.

As part of the review, I identified commonly used digital assessment formats relevant to physics instruction, including:

- Multiple-choice questions;
- Open-ended questions;
- Matching tasks (e.g., text-to-text, text-to-image, image-to-image);
- Embedded questions within videos (e.g., multiple choice, open-ended);
- Drawing or diagram creation with teacher feedback;
- Interactive tasks integrated with simulations;
- Fill-in-the-blank items (text or numerical);
- Table completion tasks (text or numerical).

To be considered suitable for use in the classroom, a tool ideally needed to support several of these assessment types. Following the initial search, the identified tools were categorized based on their core functionality (e.g., content delivery, interactivity, assessment features). Examples of these tools and their classification are presented in the Results section.

Pilot study

Based on the initial tool review, a subset of platforms was selected for a pilot implementation. Selection criteria included functionality, ease of use, and cost-effectiveness. The following tools were chosen:

- **Formative.com** and **Miro.com** for the delivery of self-paced learning materials;
- **PhET Interactive Simulations** for visualizing physics concepts through interactive simulations;
- **Wizer.me** for administering formative assessments;
- **Google Classroom** for organizing materials, communicating with students, and managing deadlines.

A unit from the existing curriculum was adapted using these tools to create a digital learning module. The module was implemented as a pilot project at Liberty High School in Hillsboro, Oregon (USA).

The pilot involved two ninth-grade classes, with a total of 50 students aged 14–15. The topic covered was *"Waves, Sound, and Sound Propagation."* During the first session, students were introduced to the topic using Miro boards to explore fundamental concepts. This was followed by two demonstrations: a hands-on activity with a slinky to illustrate mechanical wave propagation, and an interactive simulation using the "Wave on a String" tool from PhET. Students then engaged in individual and group activities using laptops. Formative.com hosted a series of digital tasks incorporating videos, simulations, and short exercises. Group work included hands-on investigations, such as examining how the thickness of a rubber band or guitar string influences vibration frequency. At the end of each session, students completed an assessment on Wizer.me. All resources, activities, and communications were managed via Google Classroom to ensure centralized access.

To evaluate the usability of the digital tools, feedback was collected through a teacher interview. This feedback provided initial insights into the effectiveness and accessibility of the selected platforms in a real classroom environment.

Workshops

Following the pilot study, the implementation strategy was streamlined by consolidating multiple tools into a single platform. Nearpod.com was selected as the primary environment for content delivery, interaction, and assessment due to its versatility and ease of integration.

To disseminate the updated curriculum and gather broader feedback, the materials were presented in teacher workshops at two international physics education conferences: GIREP 2023 (Groupe International de Recherche sur l'Enseignement de la Physique) and MPTL 2023 (Multimedia in Physics Teaching and Learning). These sessions aimed to showcase the instructional design and technological tools while exploring their applicability in diverse educational settings.

The first workshop was attended by approximately 20 participants, and the second by 6 participants. Attendees primarily included physics educators and researchers. Participants experienced a simulated physics lesson using the digital tools from both student and teacher perspectives. Activities included real-time interaction with embedded simulations and questions, as well as an overview of the teacher interface, which allows instructors to monitor student progress, respond to queries, and provide immediate feedback.

At the end of each session, participants were invited to discuss their impressions and provide written feedback through an optional online survey. Four completed surveys were submitted, offering valuable insights into the perceived usability, pedagogical potential, and possible limitations of the tools demonstrated.

**3. Results**

Tool Categories

Based on the analysis of core functionality and features, the identified digital tools were grouped into the following categories:

1. **Learning Management Systems (LMS):**
   Platforms in this category support the delivery and organization of instructional content. They allow educators to create lessons, distribute files or presentations, assign tasks, and share materials with students. Importantly, these resources are accessible outside of scheduled class time, enabling flexible, asynchronous learning.
2. **Video Conferencing Tools:**
   This category includes platforms designed for real-time communication, supporting scheduled virtual meetings and classes with multiple participants.
3. **Content Creation Tools:**
   Platforms that specialize in the development of instructional materials in specific formats (e.g., images, videos, digital books, presentations). These tools often offer templates and editing features tailored to particular media types.
4. **General Assessment Tools:**
   Tools in this category support the creation of diverse assessment formats, such as multiple-choice, open-ended questions, and fill-in-the-blank tasks. They typically allow for the integration of multimedia elements (images, video, audio) and the combination of various task types within a single worksheet.
5. **Media-Specific Assessment Tools:**
   These platforms are assessment-focused but rely predominantly on one or two media types as the primary basis for evaluation, for example, tools designed specifically for video-based quizzes or reading comprehension tasks.
6. **Collaboration Tools:**
   Designed to facilitate group work and joint content development, these tools may include features such as shared whiteboards, discussion forums, and real-time commenting. While not always education-specific, they are widely used in instructional contexts to support peer interaction and project-based learning.

7. **Quiz Platforms:**

    This category includes software dedicated to creating and administering online quizzes. Unlike general assessment tools, these platforms are typically used synchronously during class sessions and often support group discussion and immediate feedback.

8. **Specialized Media and Simulation Tools:**

    These tools focus on the creation or deployment of domain-specific content, such as scientific simulations or interactive models. They often require subject-specific knowledge and technical expertise to use effectively and are particularly valuable in STEM education.

Examples of representative tools for each category are summarized in Table 1.

Table 1 - Online resources examples based on outlined categories

| **Learning Management Systems:** | **Video Conferencing Tools:** |
| --- | --- |
| Nearpod | Google Meet |
| Classkick | Zoom |
| Kami | Teams |
| Pear Deck | |
| Moodle | |
| Google Classroom | |
| **General Assessment Tools:** | **Content Creation Tools:** |
| Formative | Book creator |
| Wizer.me | Canva Education |
| | Genially |
| | Flip (ex. Flipgrid) |
| | Iorad |
| | AI based tools |
| **Collaboration Tools:** | **Media-Specific Assessment Tools:** |
| Wakelet<br>Miro.com<br>Jamboard<br>Padlet | Activelylearn<br>Edpuzzle<br>Classcraft |
| **Specialized Media and Simulation Tools:**<br>PhET simulations<br>GeoGebra<br>Wolfram Alpha | **Quiz Platforms:**<br>Learningapps<br>Quizlet<br>Mentimeter<br>Wordwall<br>Kahoot<br>Gimkit |

The list of tools presented in this study is not intended to be exhaustive. Given the continuous emergence of new educational technologies, the selection reflects those platforms considered most relevant at the time of the research.

Following the initial pilot implementation, five educational tools, **Miro.com**, **Formative.com**, **Wizer.me**, **PhET simulations**, and **Google Classroom**, were tested in classroom settings. Each tool offered distinct advantages, but also presented limitations that informed subsequent adjustments in tool selection.

**Miro.com** offered a highly collaborative environment with an infinite whiteboard interface suitable for teamwork and brainstorming. Figure 1a shows an overview of the developed whiteboard. However, teachers reported significant technical difficulties including long loading times, connection instability, and a non-intuitive interface, especially for first-time users without touchscreen devices. These usability issues outweighed the collaborative potential in practice.

**Wizer.me** enabled the creation of interactive worksheets using visually appealing templates and access to a large library of shared resources. An example of an assessment question is shown in figure 1b. Yet, the necessity to publish resources publicly in the free version and the limited types of questions and analytics available posed challenges in classroom implementation.

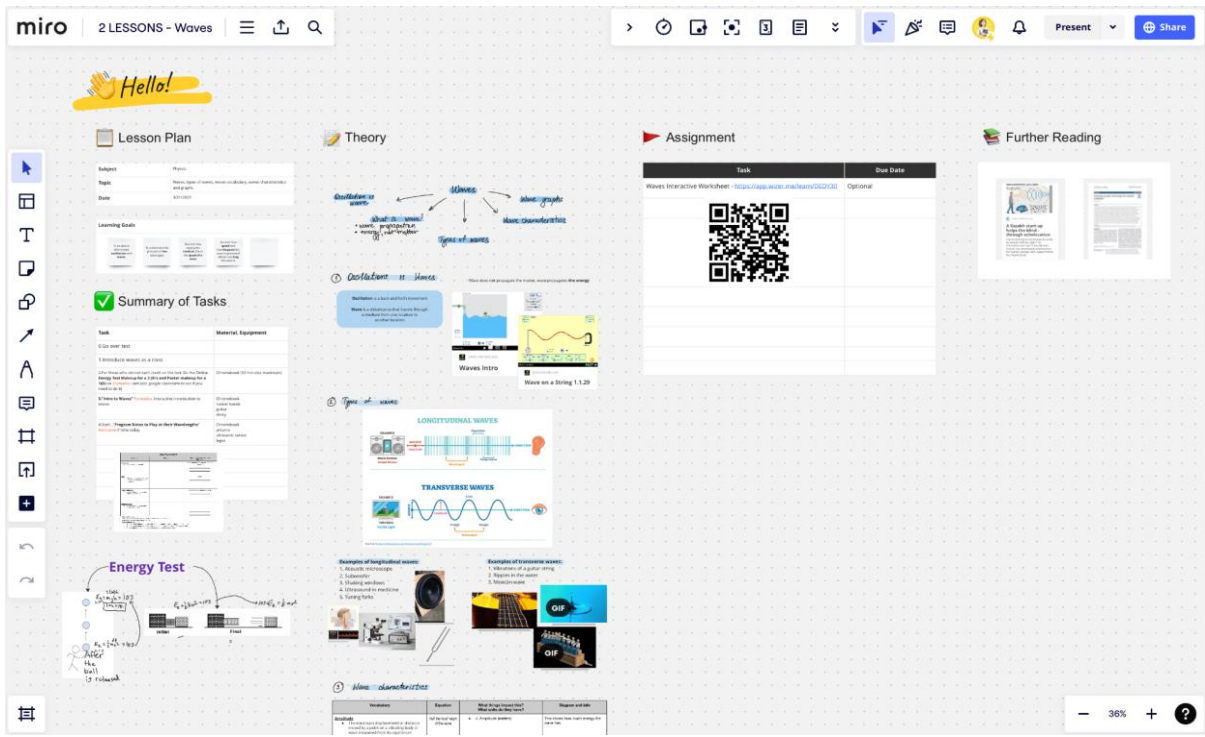

a)

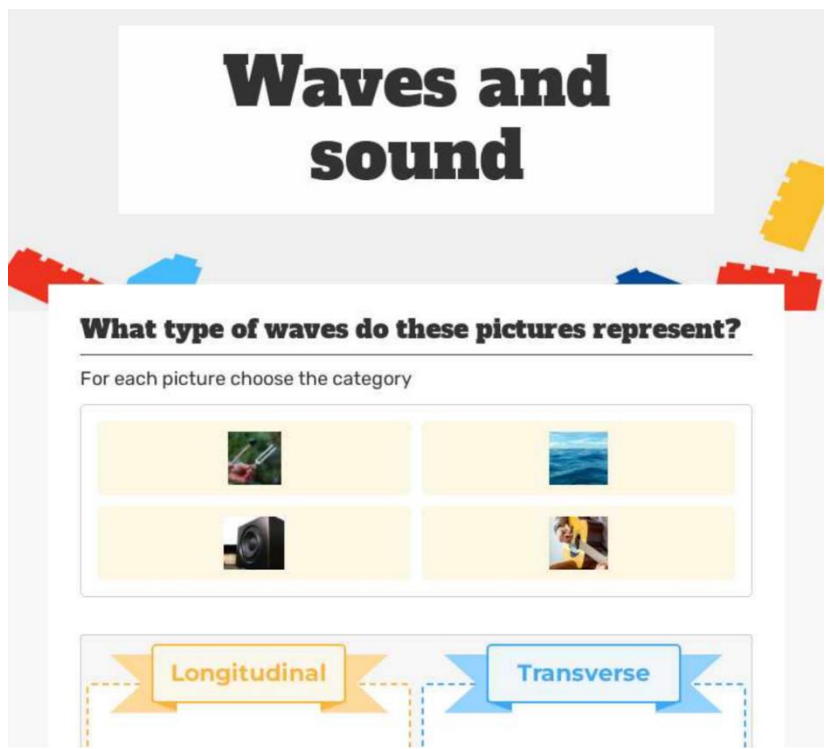

b)

Figure 1 - a) Miro board created for waves topic, b) preview of Wizer worksheet

**Formative.com** stood out for its real-time analytics, diverse question types, and seamless integration with Google Classroom. However, the limited visual customization and styling of worksheets often made it difficult to emphasize key information, resulting in students skipping important content unintentionally. Figure 2 shows an example question of the final worksheet for students.

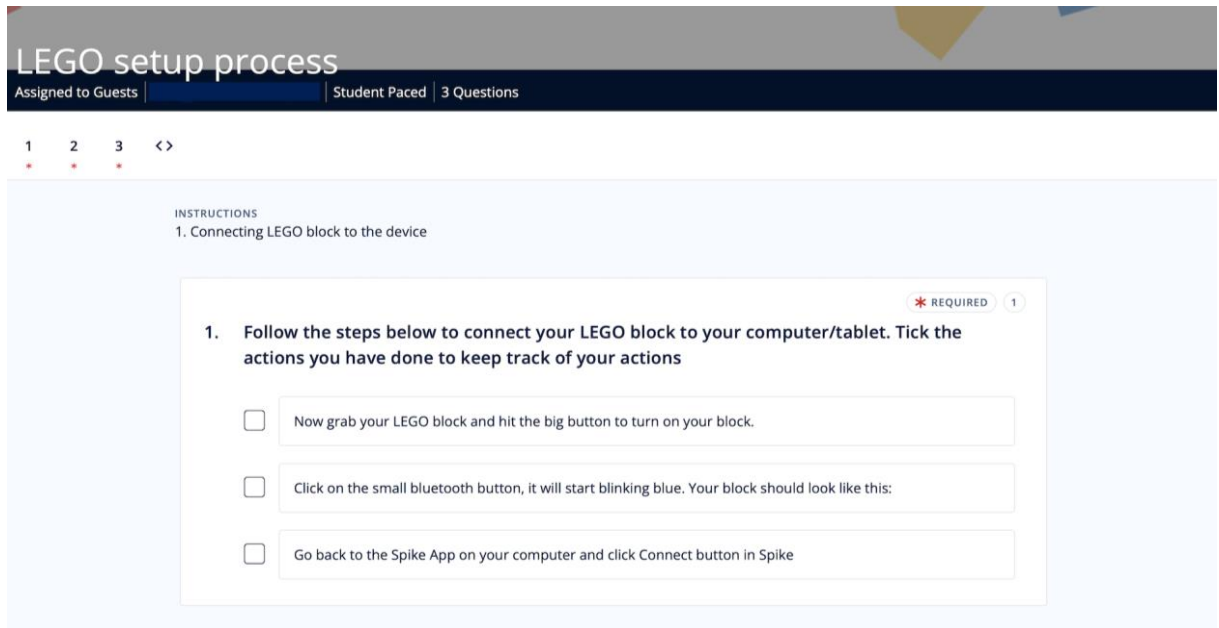

Figure 2 - Formative worksheet overview,

While each of these platforms addressed specific classroom needs, simultaneous use of multiple platforms resulted in a fragmented experience. The switching between tools was found to be confusing for both students and teachers and reduced the overall effectiveness of the digital lesson flow.

In response, subsequent workshops adopted **Nearpod.com** as a unified solution. The developed set of labs in the Nearpod interface is shown in figure 3. Nearpod combines presentation features with built-in assessments, real-time feedback, and student-paced modes. Teachers highlighted its accessibility without user registration, its visual presentation style that reduced skipped content, and its capacity to integrate multiple learning activities into one coherent experience. Drawbacks included limited media integration (some content opened in separate windows) and the difficulty of processing post-lesson data in numerical form.

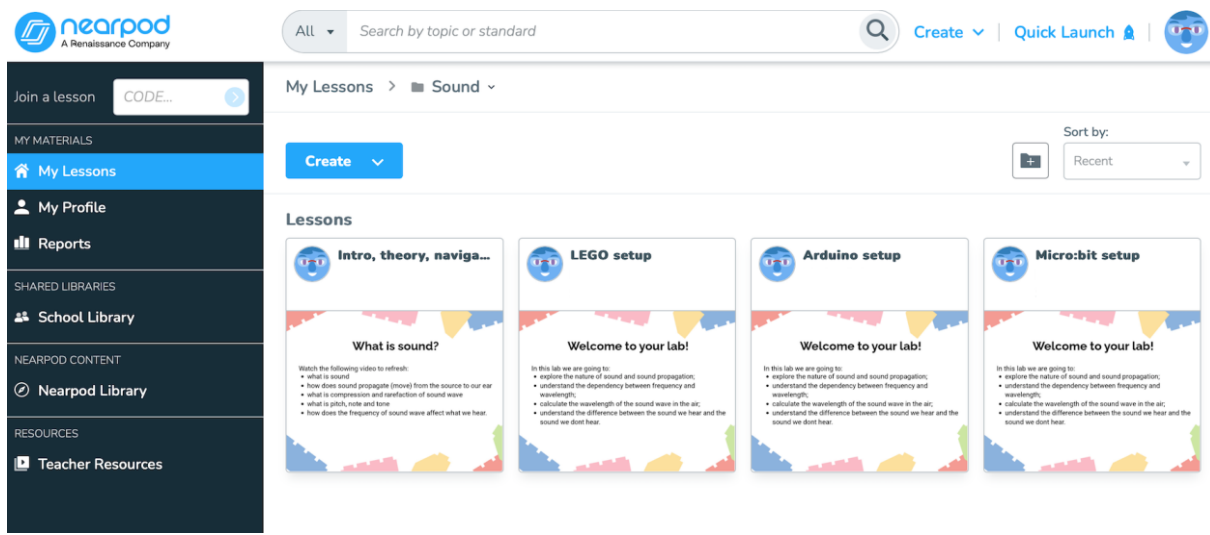

Figure 3 - Nearpod lessons library

Workshop feedback and teacher interviews supported these findings. Teachers across all levels found collaboration, content sharing, and tool accessibility to be the most valuable aspects. Reported challenges included internet connectivity issues, steep learning curves for some tools, and the need for clearer student instructions. Several teachers also highlighted the lack of inclusive materials for learners with special needs and requested more targeted resources and support.

Despite these challenges, nearly all respondents reported a high likelihood of incorporating the tested tools into future teaching. Additionally, several educators expressed interest in receiving further training, both self-paced and instructor-led, on the use of collaborative and digital tools.

## 4. Discussion

The findings from the pilot study and subsequent educator workshops demonstrate the significant potential of digital tools to enhance physics instruction, particularly when thoughtfully integrated into lesson design. However, the implementation also revealed practical challenges that must be addressed for these tools to be effective in real-world classrooms. A key takeaway from this study is the value of platform consolidation. While tools such as Miro, Wizer, and Formative each supported specific instructional functions, collaboration, formative assessment, and real-time feedback, their simultaneous use often resulted in fragmented lesson structures. This fragmentation led to cognitive overload for students and disrupted instructional flow, particularly in younger learners. These findings echo the concerns raised by Skulmowski and Xu (2021).

In contrast, Nearpod, which was adopted for the workshops, provided a more streamlined and coherent digital experience. Its integrated features, such as interactive slides, embedded assessments, and real-time monitoring, enabled teachers to deliver lessons in a unified format, reducing the need for transitions between platforms. Participants appreciated its visual clarity, intuitive interface, and accessibility, especially the ability to join sessions without student accounts. These characteristics align with recommendations in the literature advocating for tools that support blended and self-paced learning environments (Guo et al., 2023; Xu et al., 2023). Nevertheless, even Nearpod presented limitations. For instance, it has a restricted media integration (e.g., certain content opening in separate windows) and its limited capacity for exporting and analyzing detailed assessment data. These issues underscore a broader challenge in educational technology: balancing usability with feature richness. Tools that are simple and easy to adopt may lack the flexibility needed for differentiated instruction, while feature-rich tools may require substantial training and support to use effectively.

Teacher and workshop participant feedback emphasized the importance of collaboration, access, and adaptability. Tools that supported easy content sharing and peer-to-peer interaction were rated highest, especially in environments where infrastructure (such as stable internet connections or available hardware) could not be guaranteed. The importance of accessibility and inclusivity also emerged as a recurring theme in teacher feedback. Educators highlighted the need for materials that could accommodate learners with physical or sensory impairments.

It is also noteworthy that educators reported changes in classroom dynamics when new tools were introduced. Some observed increased engagement from students who were typically less active, suggesting that digital tools can reach learners who may not respond to traditional teaching methods. Others highlighted the novelty of hands-on materials as a motivating factor, especially when paired with digital instructions or collaborative tasks.

Looking ahead, the findings point toward an emerging challenge: the integration of artificial intelligence (AI) in physics education. While AI tools hold promises for personalized learning, adaptive feedback, and intelligent content delivery, they also raise new concerns around transparency, teacher readiness, and ethical use. Effective AI deployment will depend on robust data analytics and careful interpretation of learner behavior, capacities that most current platforms do not yet support at scale.

Ultimately, the findings highlight a need for teacher-centered implementation strategies, where the selection and use of tools are guided by classroom context, infrastructure limitations, and learner diversity. The study reinforces the idea that technology alone does not immediately improve education; it must be aligned with pedagogical goals, adapted to the teaching context, and supported by adequate training and infrastructure. Even well-designed tools require support structures such as clear instructions, technical troubleshooting, and targeted training to be fully effective. Future research should explore long-term implementation across diverse learning environments and evaluate the impact of newer technologies, including AI-driven systems, on student learning outcomes and teacher practices.

## 5. Limitations

This study was exploratory in nature and subject to several limitations. First, the pilot implementation and workshops were conducted with a small sample of educators, which may not fully represent the diversity of teaching contexts, school infrastructures, and student populations. Therefore, the generalizability of the findings is limited.

Second, the duration of the tool implementation was relatively short. Teachers and students may require more extended use to fully adapt to the tools and provide deeper insights into their long-term impact on learning and engagement.

Third, the effectiveness of the tools was influenced by external factors such as internet connectivity, hardware availability, and individual teachers' prior experience with educational technology. These factors varied across participants and could have skewed perceptions of tool usability and usefulness.

Finally, while feedback was collected from both interviews and written reflections, it remained largely qualitative and self-reported. A more systematic approach, including quantitative performance measures and classroom observations, would provide a more robust understanding of the tools' impact.

## 6. Conclusion

Successful implementation of online learning tools and management software can support students' learning processes, increase accessibility, and promote inclusiveness in education. This paper presented an overview and categorization of available digital tools that can assist educators in creating scaffolding, organizing content, conducting assessments, and fostering collaborative learning environments. The aim was to identify practical and accessible tools that could enrich lesson delivery, enhance student engagement, and provide teachers with flexible instructional strategies. The developed categories can serve as a valuable reference for educators exploring new approaches in their practice.

Looking ahead, the integration of AI into physics education represents a rapidly evolving frontier. AI offers promising applications, from intelligent tutoring systems and adaptive assessments to automated feedback and data-driven personalization. However, it also presents new challenges related to pedagogical design, equity, transparency, and teacher readiness. As educational technology continues to advance, supporting educators in understanding and thoughtfully implementing AI will be a crucial next step in the journey toward more effective and future-ready physics instruction.

## Acknowledgement

The work was supported by the grant SVV no. 260828.